\documentclass[sigconf]{acmart}

\AtBeginDocument{%
  }

\copyrightyear{2023}
\acmYear{2023}
\setcopyright{acmlicensed}\acmConference[SIGIR-AP '23]{Annual International ACM SIGIR Conference on Research and Development in Information Retrieval in the Asia Pacific Region}{November 26--28, 2023}{Beijing, China}
\acmBooktitle{Annual International ACM SIGIR Conference on Research and Development in Information Retrieval in the Asia Pacific Region (SIGIR-AP '23), November 26--28, 2023, Beijing, China}
\acmPrice{15.00}
\acmDOI{10.1145/3624918.3625311}
\acmISBN{979-8-4007-0408-6/23/11}

\usepackage{hyperref}
\usepackage{latexsym}
\usepackage{mathrsfs}
\usepackage{multirow}
\usepackage{amsmath}
\usepackage{amsfonts}
\usepackage{subfigure}
\usepackage{graphicx}
\usepackage{algorithmic}
\usepackage{comment}
\usepackage{newtxmath}
\usepackage{balance}
\usepackage{multirow}
\usepackage{algorithm}

\begin{document}

\title{Reinforcement Re-ranking with 2D Grid-based \\ Recommendation Panels}

\author{Sirui Chen}
\affiliation{
    \institution{School of Information \\Renmin University of China}
    \city{Beijing}
    \country{China}
    \\ chensr16@gmail.com
}

\author{Xiao Zhang}
\affiliation{
    \institution{Gaoling School of Artificial Intelligence \\Renmin University of China}
    \city{Beijing}
    \country{China}
    \\ zhangx89@ruc.edu.cn
}
\author{Xu Chen}
\affiliation{
    \institution{Gaoling School of Artificial Intelligence \\Renmin University of China}
    \city{Beijing}
    \country{China}
    \\ successcx@gmail.com
}

\author{Zhiyu Li}
\affiliation{
    \institution{Alibaba Group}
    \city{Hangzhou}
    \country{China}
\ tuanyu.lzy@taobao.com
}

\author{Yuan Wang}
\affiliation{
    \institution{Alibaba Group}
    \city{Hangzhou}
    \country{China}
\ wy175696@taobao.com
}

\author{Quan Lin}
\affiliation{
    \institution{Alibaba Group}
    \city{Hangzhou}
    \country{China}
\ tieyi.lq@taobao.com
}

\author{Jun Xu}
\authornote{Corresponding author. Work partially done at Engineering Research Center of Next-Generation Intelligent Search and Recommendation, Ministry of Education.}
\affiliation{
    \institution{Gaoling School of Artificial Intelligence \\Renmin University of China}
    \city{Beijing}
    \country{China}
\\ junxu@ruc.edu.cn
}

\renewcommand{\shortauthors}{Sirui Chen et al.}

\begin{abstract}
Modern recommender systems usually present items as a streaming, one-dimensional ranking list. 
Recently there is a trend in e-commerce that the recommended items are organized grid-based panels with two dimensions where 
users can view the items in both vertical and horizontal directions. 
Presenting items in grid-based result panels poses new challenges to recommender systems because existing models are all designed to output sequential lists while the slots in a grid-based panel have no explicit order. Directly converting the item rankings into grids (e.g., pre-defining an order on the slots) overlooks the user-specific behavioral patterns on grid-based panels and inevitably hurts the user experiences. 
To address this issue, we propose a novel Markov decision process (MDP) to place the items in 2D grid-based result panels at the final re-ranking stage of the recommender systems. The model, referred to as Panel-MDP, takes an initial item ranking from the early stages as the input. Then, it defines \emph{the MDP discrete time steps as the ranks in the initial ranking list, and the actions as the prediction of the user-item preference and the selection of the slots}. At each time step, Panel-MDP sequentially executes two sub-actions: first deciding whether the current item in the initial ranking list is preferred by the user; then selecting a slot for placing the item if preferred, or skipping the item otherwise. The process is continued until all of the panel slots are filled.
The reinforcement learning algorithm of PPO is employed to implement and learn the parameters in the Panel-MDP. 
Simulation and experiments on a dataset collected from a widely-used e-commerce app demonstrated the superiority of Panel-MDP in terms of recommending 2D grid-based result panels. 
\end{abstract}

\ccsdesc[500]{Information systems~Recommender systems}
\ccsdesc[500]{Computing methodologies~Reinforcement learning}

\keywords{re-ranking, reinforcement learning, recommender system}


\maketitle

\section{Introduction}
In recent years, there is a trend in e-commerce recommendation apps that the results can be presented in 2D grid panels so that the users can browse results vertically and horizontally. Studies have shown that comparing with the traditional 1D ranking list, 2D grid panels are more attractive since the users can make broader and deeper explorations~\cite{kammerer2010interface}. 

Unlike 1D ranking list where the users' top-down browsing manner makes it possible to directly allocate the items from the top according to their ranked preference scores, in 2D panels, the user feedback is gained after the user browses the whole page. This makes that the correspondences between the slots and the preference scores are not immediate, which is usually overlooked in traditional recommender systems.
One may consider straightforward heuristics, such as expanding the panels from the top-left to the bottom-right, to degenerate the 2D re-ranking problem into a conventional 1D re-ranking task.
However, the user interaction patterns on grid panels are different from that of the ranking list~\cite{xie2019grid}: they browse the result panel in both horizontal and vertical directions; their attentions are biased to the middle positions, and their attentions do not decrease monotonically and dramatically with the ranks. Directly mapping the slots in a 2D grid-based panel to the ranks in a 1D list deviates from the users' activity patterns and inevitably hurts the user experiences. 

To address the issue, this paper proposes a novel Markov decision process (MDP)-based re-ranking model for final-stage recommendation, called Panel-MDP. Specifically, Panel-MDP formulates the process of pruning and placing the items from an initial ranking list to a grid-based result panel as sequential decision making. To overcome the difficulty that the slots in a grid-based panel have no explicit order, unlike existing MDP-based re-ranking models~\cite{bello2018seq2slate, xu2020reinforcement, zhao2018deep}, Panel-MDP defines its time steps as the ranks in the initial list. Its actions are defined as compositions of two parts, respectively correspond to the two factors in the final re-ranking stage: first judging whether to expose the current item and then selecting the position to place the item if it needs to be exposed. 

With the above configuration, Panel-MDP is enabled to generate grid-based results naturally. At each time step $t = 0, 1, \cdots$, Panel-MDP processes the $(t+1)$-th item in the initial ranking list: according to the current state and guided by the policy, it first selects an action $a = (a^\textrm{exp}, a^\textrm{pos})$, where $a^\textrm{exp}$ denotes the selection of the item, and $a^\textrm{pos}$ denotes the position to be placed. The item is placed to the slot indexed by $a^\textrm{pos}$ if $a^\textrm{exp} = 1$, otherwise the item is discarded. Then, the MDP state is updated for processing the next item in the initial ranking. The process will continue until the result panel is full or all items in the initial ranking have been processed. To learn the model parameters, the reinforcement learning algorithm of Proximal Policy Optimization (PPO) is employed, which takes the (simulated) user feedback on the grid-based result panels as the training signals. 

Panel-MDP is tailored for 2D grid-based result panels, and offers several advantages: (1) the novel configuration of time steps and actions enables it to generate grid-based result panels naturally. Moreover, it has the potential to generate more sophisticated result layouts; (2) it enables the model to capture the special user activity patterns on grid-based result panels during the model learning.

In summary, the main contributions of this paper are presented as follows:
(1) we highlight a new re-ranking task where the items are allocated into two-dimensional grid-based panels; (2) we propose a novel MDP-based final-stage ranking model, called Panel-MDP, to tackle the task; (3) we demonstrated the effectiveness of Panel-MDP against the state-of-art baselines on a real-world dataset collected from a popular e-commerce app.

\section{Preliminaries}
Suppose that a user $u\in\mathcal{U}$ accesses a recommender system and the early-stage of the recommendation pipeline retrieves a set of $K$ items and organizes them as an initial ranking list: $\mathbf{L}=[i_1,\cdots,i_K]\in \mathcal{I}^K$ where $\mathcal{I}$ is the set of all items, and $i_k\in\mathcal{I}$ is the $k$-th item in $\mathbf{L}$. Items are typically represented as real-valued embeddings, denoted by 
$\mathbf{e}(i) \in \mathbb{R}^{d}$ where $d$ is the dimension of the item embeddings.

In the final-stage recommendation, (a subset of) these $K$ items are placed to a grid-based panel of size $M$ rows and $N$ columns, denoted as $\mathbf{P} \in \mathcal{I}^{M\times N}$, where each position $P_{m,n} = i_k$ means item $i_k$ is placed to the $(m, n)$-th position in the panel. Note that usually $M\times N \leq K$. After the user $u$ has viewed a recommended panel, the system records the user feedback (e.g., 1 for purchase and 0 otherwise) which can be considered as the labels or rewards for MDP. The goal of the final-stage recommendation is to output a grid-based panel $\mathbf{P}$ based on the input ranking list $\mathbf{L}$.

One straightforward approach to placing the items in grids is pre-defining a total order by using some heuristics (e.g., ordering the positions in a panel from left to right, and then top to down). Therefore, the task de-generates to traditional re-ranking in recommendation and any re-ranking models can be directly applied. Though preliminary successes have been achieved~\cite{bello2018seq2slate}, the previous studies have also shown that user behaviors on 2D grid-based panels are very different from those on traditional sequential list-based rankings~\cite{xie2019grid}. First of all, since users can browse items vertically and horizontally, they will allocate more attention to the middle part of the panel, called middle bias; Second, user studies also showed that the users' attention decay slowly by row and they often display row-skipping behavior; Third, we found that on product apps, the users tend to click one item on a result panel though they may have interest on multiple items. Note that though some of the above conclusions were drawn based on image search, we also observe similar phenomena in e-commerce recommendation.

The analysis indicates that directly applying the traditional re-ranking models to generate grid-based result panels is not an optimal solution. It is necessary to design a new model tailored for the task. 

\section{Our Approach: Panel-MDP}
\subsection{Configuration of Panel-MDP}\label{sec:Configuration}
Existing MDP-based re-ranking models usually define the time steps to the ranks in the output result list, and the actions to the input items. This configuration, however, can only generate ranking lists because the time steps naturally form a sequence.

Different from existing MDP-based ranking models, Panel-MDP defines its \emph{time-steps as the ranks in the input list~$\mathbf{L}$, and the actions are defined as the user-item preference predictions and the slots in the panel $\mathbf{P}$}. 
In this way, Panel-MDP naturally generates a grid-based result panel, through processing the items $\mathbf{L}$ in a sequential manner.      

Specifically, Panel-MDP's time step, state, action, state transition, reward function and discount factor are defined as follows:

\textbf{Time step} $t \in \{0, \cdots, K-1\}$: We define the time step $t$ on the input ranking list $\mathbf{L}=[i_1\cdots, i_K]$. Therefore, for $\mathbf{L}$ with $K$ items, the Panel-MDP will run at most $K$ steps, each processing the item ranked at $(t+1)$-th position in $\mathbf{L}$, denoted $i_{t+1}$.

\textbf{Action} $a = (a^\textrm{exp}, a^\textrm{pos}) \in \mathcal{A} \subseteq \mathcal{A}^\textrm{exp}\times \mathcal{A}^\textrm{pos}$, where the sub-action $a^\textrm{exp}\in \mathcal{A}^\textrm{exp}=\{0,1\}$ indicates whether the current item is selected to expose: 1 means expose the item and 0 otherwise. The sub-action $a^\textrm{pos}\in \mathcal{A}^\textrm{pos}\subseteq\{(m,n)|m = 1, \cdots, M; n = 1,\cdots,N\}$ indicates the index of the slot in the grid-based result panel. 

Specifically, at the $t$-th time step, the agent of Panel-MDP processes the item $i_{t+1}$ in $\mathbf{L}$ and selects an action $a_t =(a^\textrm{exp}_t, a^\textrm{pos}_t=(m,n))$. The agent will discard item $i_{t+1}$ if $a^\textrm{exp}_t = 0$, without considering the values of $a^\textrm{pos}_t$. Otherwise (i.e., $a^\textrm{exp}_t=1$), item $i_{t+1}$ will be placed to the $(m,n)$-th slot in panel matrix $\mathbf{P}$, i.e., $P_{a^\textrm{pos}_t} = P_{m,n} \leftarrow i_{t+1}$.
Note that one slot at the result panel can be placed at most one item. The slot index will be removed from $\mathcal{A}^\textrm{pos}$ if it has been used. Formally, let's denote the action space at the $t$-th step as $\mathcal{A}_t = \mathcal{A}^\textrm{exp}_t \times \mathcal{A}_t^\textrm{pos}$. After choosing an action $a_t=(a^\textrm{exp}_t, a^\textrm{pos}_t)$ and moving to the time step $(t+1)$, the action space becomes $\mathcal{A}_{t+1} = \mathcal{A}^\textrm{exp}_{t+1}\times \mathcal{A}^\textrm{pos}_{t+1}$ where 
\begin{equation}\label{eq:UpdateA}
\mathcal{A}^\textrm{pos}_{t+1} =\begin{cases}
\mathcal{A}^\textrm{pos}_{t} & a^\textrm{exp}_t = 0,\\
\mathcal{A}^\textrm{pos}_{t}\setminus \{a^\textrm{pos}_t\} & a^\textrm{exp}_t = 1,
\end{cases} 
\end{equation}
and at the initial time-step $t=0$, $\mathcal{A}^\textrm{pos}_0 = \{(m,n)|m=1,\cdots, M; n=1,\cdots, N\}$. $\mathcal{A}^\textrm{exp}_t = \{0, 1\}$ for all time-steps $t$. 

\textbf{State $s\in \mathcal{S}$:} The state of the Panel-MDP at the $t$-th time step can be described as a tuple $s_t=[\mathbf{L}_t, t, \mathbf{H}_t]$ where $\mathbf{L}_t=[i_t,\cdots,i_K]$ denotes the candidate items which have not been seen yet in the input list, and $\mathbf{H}_t=[a_0,\cdots, a_{t-1}]$ is the sequence of actions already selected until step $t$. At the time step $t = 0$, the state is initialized as $s_0 = [\mathbf{L}, 0, \mathbf{H}_0=\emptyset]$, where $\emptyset$ is the empty list. 

\textbf{State transition $T$:} After issuing the action $a_t$ at state $s_t$, the system will transit to a new state. The state transition function $T: \mathcal{S}\times \mathcal{A}\rightarrow \mathcal{S}$ is defined as:
\[
\begin{split}
s_{t+1} & = T(s_t=[\mathbf{L}_t=[i_t,\cdots,i_K], t, \mathbf{H}_t = [a_0, \cdots, a_{t-1}]], a_t)\\
&= [\mathbf{L}_{t+1}=[i_{t+1},\cdots,i_K], t+1, \mathbf{H}_{t+1} = [a_0, \cdots, a_{t-1}, a_t]],
\end{split}
\]
which simply discards the current item $i_t$ from $\mathbf{L}_t$, increases the time step by one, and append the chosen actions $a_t$ at the $t$-th step to the action history $\mathbf{H}_t$. 

\textbf{Rewards $r_t = R(s_t, a_t)$:} After presenting a whole result panel to a user, the user may give feedback on the recommended items, i.e., purchase or not. Therefore, we can define the reward based on that feedback. For example, $r_t = R(s_t, a_t = (1, (m,n))) = 1$ if the user purchases the item at the $(m,n)$-th position, and 0 otherwise. 

\textbf{Discounted Factor} $\gamma\in(0, 1]$ discounts rewards at a constant rate per time step.

\subsection{Model Structure and Learning}
We present an implementation of Panel-MDP policy with a famous Actor-Critic algorithm of Proximal Policy Optimization (PPO)~\cite{schulman2017proximal}.
\subsubsection{Model Structure}
The implementation of our Panel-MDP framework with PPO comprises three components: a state encoder to capture the representation of a given state; a policy network to make an action based on the state, and an advantage network to measure the advantage of action in the current state.

In general, the state representation can be described as a concatenation of the component embeddings in the state:
\[
    \mathbf{e}^s  = \mathbf{concat}(\mathbf{e}^L, \mathbf{e}^t, \mathbf{e}^H),
\]
where operator `$\mathbf{concat}$' concatenates all the inputted vectors, $\mathbf{e}^t$ is the embedding of the time step $t$ which is automatically learned in an end2end manner;  $\mathbf{e}^L$ encodes the candidate item list $\mathbf{L}_t=[i_t, \cdots, i_K]$, and is defined as the mean pooling of the output vectors by a multi-layer Perceptron(MLP) layer:
$
    \mathbf{e}^L = \textbf{Mean}(\textbf{MLP}_1([\mathbf{e}^i_t,$
    \\
    $\cdots,\mathbf{e}^i_K]),
$
where \textbf{Mean}($\cdot$) denotes the mean pooling of all the input vectors, and $\mathbf{e}^i_t$ is the embedding of the $t$-th item in the input list $\mathbf{L}$.
As for $\mathbf{e}^H$, it encodes the previous action sequence $\mathbf{H}_t=[a_1,\cdots,a_{t-1}]$. It is defined as the final output vector of a GRU after the whole sequence has been scanned:
$
    \mathbf{e}^H = \textbf{GRU}([\mathbf{e}^a_1, \cdots, \mathbf{e}^a_{t-1}]),
$
where $\mathbf{e}^a$ are action embeddings which can also be automatically learned in an end2end manner.

For policy network,  the policy $\pi$ for the first sub-action $a^\textrm{exp}$ is:
\begin{equation}\label{eq:policy_as}
\pi(a^\textrm{exp} = 1|s) = \sigma(\textbf{MLP}_3(\mathbf{h})),
\end{equation}
where $\sigma$ is the sigmoid function, and $\mathbf{h} = \textbf{MLP}_2(\mathbf{e}^s)$ is transformed from the state embedding $\mathbf{e}^s$. 

As for the policy $\beta$ for the second sub-action $a^\textrm{pos}$, we introduce the slot position embeddings and  $\forall p\in\mathcal{A}^\mathrm{pos}_t$:
\begin{equation}\label{eq:policy_ap}
\beta(a^\textrm{pos}=p|s)=\frac{\exp({\mathbf{h}}^{T}\mathbf{e}^p)}{\sum_{p'\in \mathcal{A}_t^\textrm{pos}} \exp({\mathbf{h}}^{T}\mathbf{e}^{p'})},
\end{equation}
where $\mathbf{e}^p$ (and $\mathbf{e}^{p'}$) is the embedding of slot $p$ (and $p'$). 

Considering the high cost of directly estimating the advantages, we employ the GAE method ~\cite{schulman2015high} to estimate the two advantages, namely ${A}(s,a^\mathrm{exp})$ and ${A}(s, a^\mathrm{pos})$, on the basis of the value function $V(s)$:
$V(s) = \textbf{MLP}_4(\mathbf{e}^s)$,
where $\textbf{MLP}_4$ outputs a scalar value.
\subsubsection{Model Learning}
To learn the model parameters (including those in the four MLPs, GRU, time-step embeddings, position embeddings, and action embeddings), a batch of $B$ trajectories $\mathcal{D} = \{\tau_b\}_{b=1}^B$ are collected, where each $\tau_b$ is generated by the following procedure: for $t=1,\cdots, K$, given a state $s_t$, the agent will sample an action $a_t=(a^\textrm{exp}_t, a^\textrm{pos}_t)$ from the two stochastic policies: $a^\textrm{exp}_t\sim \pi(a^\textrm{exp}|s)$ and $a^\textrm{pos}_t\sim \beta(a^\textrm{pos}|s)$. After executing $a_t$, the agent receives a reward $r_t$ and updates the state to $s_{t+1}$. The procedure is repeated until the $\mathcal{A}^\textrm{pos}_t=\emptyset$ or $t > K$. In this way, we collect a trajectory $\tau_b=\left\{(s_t,a_t = (a^\textrm{exp}_t,a^\textrm{pos}_t),r_t,s_{t+1})\right\}_{t=0}^{|\tau_b|-1}$. 

PPO is employed for conducting the learning process that amounts to optimizing the following loss function:
\begin{equation}\label{eq:loss}
    \mathcal{L}=\mathcal{L}^{Actor}+\lambda_1\mathcal{L}^{Critic}+\lambda_2\mathcal{L}^{Entropy}
\end{equation}
where $\lambda_1>0$ and $\lambda_2>0$ are trade-off coeffients, $\mathcal{L}^{Actor}$ is the loss from the actor:
\begin{equation*}
\small
    \begin{split}
    \mathcal{L}^{Actor}&=-\mathbb{E}_{\mathcal{D}}\left[\min\left(w^\textrm{exp}A(s,a^\textrm{exp}),\mathrm{clip}\left(w^\textrm{exp},1-\epsilon,1+\epsilon\right)A(s,a^\textrm{exp})\right)\right.\\
    +\mathbb{I}&\left.[a^\textrm{exp}=1]\cdot\min\left(w^\textrm{pos}A(s,a^\textrm{pos}),\mathrm{clip}\left(w^\textrm{pos},1-\epsilon,1+\epsilon\right)A(s,a^\textrm{pos})\right)\right],
\end{split}
\end{equation*}

where for each $\tau\in\mathcal{D}$ and for each tuple $(s, (a^\mathrm{exp}, a^\mathrm{pos}), r, s')\in \tau$,  $w^\textrm{exp}=\frac{\pi(a^\textrm{exp}|s)}{\pi_\mathrm{old}(a^\textrm{exp}|s)}$ and $ w^\textrm{pos}=\frac{\beta(a^\textrm{pos}|s)}{\beta_\mathrm{old}(a^\textrm{pos}|s)}$ denotes the probability ratios between new policies and old policies on the sub-action $a^\textrm{exp}$ and $a^\textrm{pos}$, respectively. The `clip' operator with parameter $\epsilon$ decides how far the new policy can deviate from the old policy. $\mathbb{I}[\cdot]$ is an indicator function which splits $\mathcal{L}^{Actor}$ into two parts: the first part corresponds to sub-action $a^\textrm{exp}$ and the second part corresponds to sub-action $a^\textrm{pos}$. The second part is dependent on the value of $a^\textrm{exp}$: when $a^\textrm{exp}=1$ (i.e., the item will be exposed at $a^\textrm{pos}$-th slot), the part will be counted in the total loss. Otherwise, this part will be ignored because the sub-action is actually not issued. 

As for $\mathcal{L}^{Critic}$, it is defined as the mean-squared loss:
\begin{equation*}
    \mathcal{L}^{Critic}=\mathbb{E}_\mathcal{\tau\sim \mathcal{D}}\mathbb{E}_t[(V(s_t)-R_t)^2],
\end{equation*}

where $R_t=\sum_{k=t}^{|\tau|-1} \gamma^{k-t} r_k$ is the discounted sum of future reward. 

Besides, we also apply an entropy-based regularizer on $\pi(a^\textrm{pos}|s)$ to encourage the agent exploration on the different positions:
\begin{equation*}
    \mathcal{L}^{Entropy}=-\mathbb{E}_\mathcal{D}[\beta(a^\textrm{pos}|s)\log\beta(a^\textrm{pos}|s)].
\end{equation*}

\begin{algorithm}[t]
    \caption{Training Panel-MDP with PPO}
	\label{alg:offline training}
	\begin{algorithmic}[1]
	\REQUIRE Initialized Actor $\pi, \beta$, Critic $V$
	\REPEAT 
	    \STATE $\mathcal{D}\leftarrow \emptyset$
	    \FOR{$b=0,\cdots,B$}
	        \STATE Initialize: $s \leftarrow[\mathbf{L}, 0, \emptyset]$; $\mathcal{A}^\textrm{exp}\leftarrow \{0, 1\}; \mathcal{A}^\textrm{pos}\leftarrow \{(m,n)|m=1,\cdots, M; n=1,\cdots,N\}$; $\tau_b \leftarrow \emptyset$
	        
	        \FOR{$t=0,1, \cdots$}
	            \STATE Sample sub-actions $a^\textrm{exp}\sim\pi,a^\textrm{pos}\sim\beta$~\COMMENT{Eq.~\eqref{eq:policy_as}\&~\eqref{eq:policy_ap}}
	            \STATE Execute $a = (a^\textrm{exp}, a^\textrm{pos})$, observe reward $r$, and new state $s'$~\COMMENT{ Sec. 3.1}
	            \STATE Update $\mathcal{A}^\textrm{pos}$~\COMMENT{Eq.~\eqref{eq:UpdateA}}
	             \STATE Append $\left(s, (a^\textrm{exp},a^\textrm{pos}),r,s'\right)$ to $\tau_b$
	        \ENDFOR
	       \STATE $\mathcal{D}\leftarrow \mathcal{D}\bigcup \{\tau_b\}$
	    \ENDFOR
	    \FORALL{$\tau\in\mathcal{D}$ and $0\leq t < |\tau|$}
	        \STATE Compute rewards-to-go $R_t$'s
	        \STATE Compute advantages $A(s, a^\textrm{exp})$ and ${A}(s, {a^\textrm{pos}})$ using GAE method~\cite{schulman2015high}, based on current value function $V$ 
	    \ENDFOR
	    \STATE Update Actor-Critic by applying SGD w/ Adam to Eq.~\eqref{eq:loss}
	\UNTIL{convergence}
	\RETURN $\pi, \beta, V$
	\end{algorithmic}
\end{algorithm}

Standard PPO-clip with gradient descent~\cite{schulman2017proximal} is applied to optimize the parameters. 
In the online re-ranking phase, after receiving the initial ranking list $\mathbf{L}$ and state initialization, Panel-MDP constructs a grid-based result panel by repeatedly executing actions through maximizing the learned policies $\pi$ and $\beta$, updating the state and actions space, and moving to the next iteration. This procedure ends until the result panel is full or all items are processed. 
\section{Experiments}
We conducted experiments to test our Panel-MDP.

To the best of our knowledge, there is no publicly available model specially designed for recommending grid-based results in the literature. Therefore, we adapted existing re-ranking models to generate grids, by simply placing the outputted item list in a predefined order(left-to-right and top-to-down). The chosen re-ranking models include the supervised model \textbf{miDNN}~\cite{zhuang2018globally}, and RL-based re-ranking model of \textbf{DDPG}~\cite{lillicrap2015continuous} and \textbf{PDQN}~\cite{zou2020pseudo}.

\textbf{Experiments on simulated environment:}
\begin{figure}[t]
    \centering

    \includegraphics[width=0.4\textwidth]{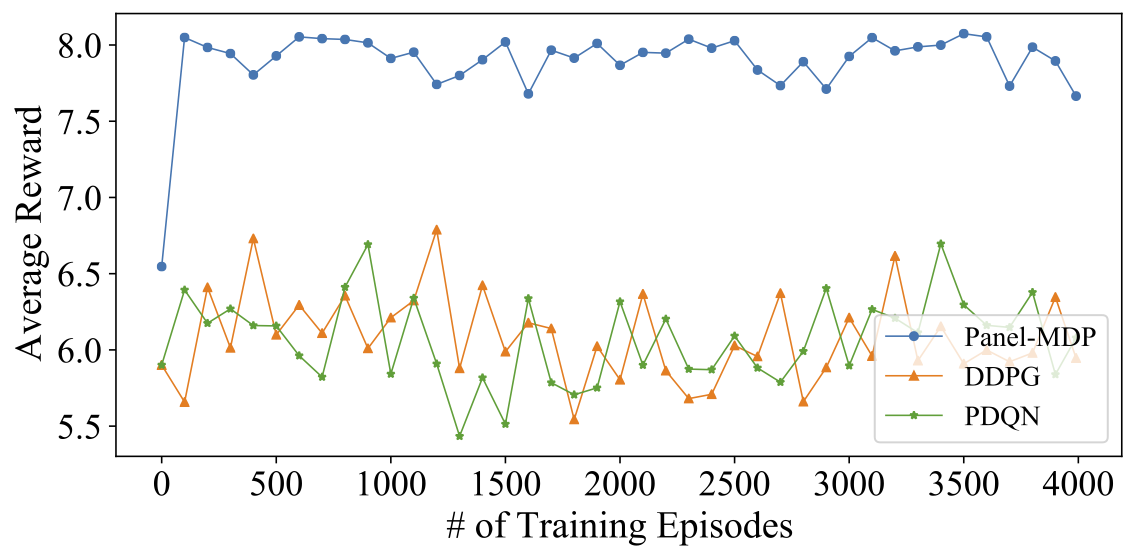}
    \caption{Performance curves of Panel-MDP and the RL-based baselines on simulated environment.}
    \label{fig:PerfCurveSimulation}

\end{figure}
We constructed a simulation environment to test whether Panel-MDP and other RL-based algorithms can capture the user activity patterns. In the simulated environment and at every session, the agent will receive a user and a ranking of 20 candidate items (represented as features). The goal is to re-rank the items in a $4\times 3$ grid. The user will provide feedback to each item based on the given grid in terms of his preference both on items and positions.  
Figure~\ref{fig:PerfCurveSimulation} demonstrates the performance curves of Panel-MDP and the baselines of DDPG and PDQN, in terms of cumulative average reward w.r.t. number of training episodes. We can observe that Panel-MDP achieves highest cumulative average reward at very beginning of the training episodes, verified the effectiveness of Panel-MDP for capturing the special user activity patterns on grid-based result panel.

\textbf{Experiments on real dataset:}
We also conducted experiments on a dataset collected from a popular e-commerce app in Oct. 2021, where the recommendation results are displayed as $2\times 3$ grid-based panels. For each user request, the early stages provided 16 ranked candidate items. The users' feedback (i.e., purchase) was collected as the labels. Users' profile (e.g., hashed user id, non-sensitive user feature etc.) and the items' profile (e.g., category, price, etc.) were stored for calculating the user and item embeddings. After pre-processing, the dataset contains 20,000 user requests for training and 1,000 for testing. 
\begin{table}[htb]
  \caption{Performance comparison between Panel-MDP and the baselines in terms of averaged reward and AUC. `$*$' means the improvements over the best baseline are statistical significant (t-tests and $p$-value $< 0.05$).}\label{tab:result}

  \label{tab:result}
  \begin{tabular}{l|c|cc}
    \toprule
    \multirow{2}{*}{Model} & \multicolumn{1}{c}{Task 1: Re-org} & \multicolumn{2}{|c}{Task 2: Select\&Re-org} \\
    \cmidrule(lr){2-2} \cmidrule(lr){3-4} 
     & Average Reward & Average Reward & AUC \\
    \midrule
    miDNN     & 0.0177               & 0.0178   & 0.7524            \\
    DDPG      & 0.0175               & 0.0145   & 0.6613             \\
    PDQN      & 0.0177               & 0.0163   & 0.7524            \\
    \midrule
    Panel-MDP (Ours) & \textbf{0.0179}$^*$        & \textbf{0.0179}$^*$   & \textbf{0.7980}$^*$          \\
    \bottomrule
  \end{tabular}

\end{table}
\begin{figure}[htb]
    \centering

\includegraphics[width=0.48\textwidth]{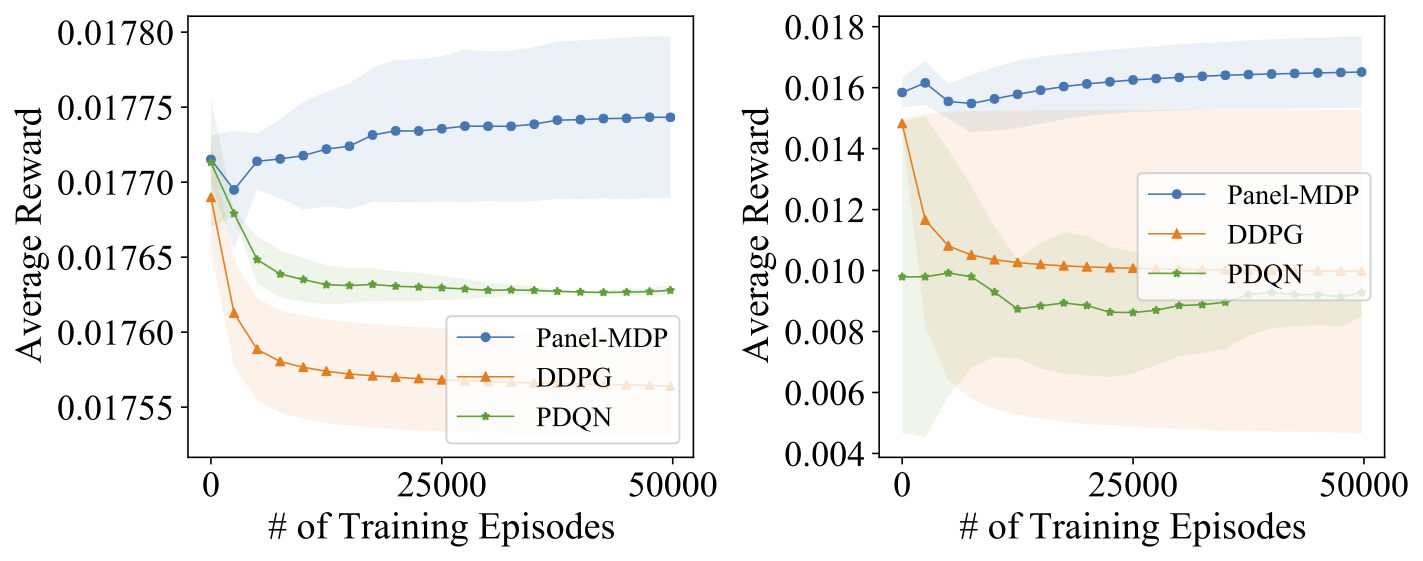}
    \caption{Performance curves of Panel-MDP and the RL-based baselines on real dataset. Left: Re-org; Right: Select\&Re-org. The shading areas represent the standard deviations under running episodes.}
    \label{fig:PerfCurve}

\end{figure}
We tested Panel-MDP on two tasks. The first task (called \textbf{Re-Org}) assumes that the early stages only provide 6 ranked items (the top 6) and the model re-organizes them to a $2\times 3$ grid, by setting sub-action $a^\textrm{exp}_t = 1$ for all $t$). The second task (called \textbf{Select\&Re-Org}) assumes that the early stages provide 16 ranked items and the model needs to select 6 items and re-organize them to a $2\times 3$ grid. As for evaluation, we use a user simulator trained on large-scale user-system interactions. Given a user, the context, and the result panel, the simulator outputs a \emph{score} denoting the purchase probability that is treated as the \emph{reward} from environments, and the average reward is reported. For Select\&Re-Org, we also used AUC for evaluation.\footnote{In the Re-Org task, AUC cannot be calculated because all the candidate items were treated positive(i.e., there is no score in RL recommender systems).}

Table~\ref{tab:result} reports the best performance in terms of the average reward on the test set. From the results, we can see that Panel-MDP outperformed all the baselines on both of the tasks, indicating the effectiveness of Panel-MDP in re-organizing the items to 2D grid-based result panels. In the task of Select\&Re-org, the action $a^\textrm{exp}$ is the key for exploration on selection. Compared to other RL baselines that have shown a decline in performance on this task, exploration allows Panel-MDP to achieve a better average reward and AUC.

Figure~\ref{fig:PerfCurve} illustrates the performance curves of Panel-MDP and the RL-based baselines in terms of cumulative average reward w.r.t. number of training episodes on the test set. Panel-MDP consistently outperformed the baselines of DDPG and PDQN, especially in the task of Select\&Re-org. This is because both DDPG and PDQN lack the ability to capture the correspondence between positions. Their performances decline fast, indicating that the configuration of Panel-MDP is suitable for generating 2D result panels. 

\section{Related Work}
Re-ranking considers the potential impact of the interrelationship of the candidate items on user preference~\cite{verstrepen2015top,jiang2017learning,wang2019sequential}.  \citet{bello2018seq2slate}~considered the influence on the previous recommended items into final-stage re-ranking models; \citet{zhuang2018globally}~proposed to fully capture the mutual information of the items in candidate sets. \citet{pei2019personalized}~proposed to further extract information in the recommended list using transformer sturcture.

Recently, reinforcement learning (RL) methods have been widely employed in recommendation to enhance long-term engagement or boost various objectives~\cite{liu2023generative, ge2021towards, xue2023prefrec, chen2023controllable}. \citet{shani2005mdp}~first applied MDP to recommendation. \citet{cai2018reinforcement}~defined a general framework for RL recommender systems. \citet{huang2021deep,zou2019reinforcement}~both utilized RL framework to model the change of users' long-term engagement and preference. \citet{zhao2018deep} proposed a page-wise RL-based recommendation model.

Compared to the re-ranking, much less research efforts have been spent on presenting results other than lists, e.g., grid-based result panels. \citet{xie2017investigating} studied the characteristics of user attentions on a 2D image search result page, and found different behavioral patterns like ``middle bias'', ``row skipping'' and ``slow decay''. \citet{guo2020debiasing}  propose a de-biasing method on 2D result pages.

\section{Conclusion and Future Work} 
This paper proposes a novel MDP-based model that formalizes the generation of a grid-based result panel from a list of ranked items as a discrete-time sequential decision making, where the time steps and actions are respectively defined as the positions in the input list and slots in the result panel. The model, called Panel-MDP, employs PPO to implement the Actor-Critic and optimize the model based on users' feedback. Experimental results on a real-world dataset collected from a popular e-commerce app showed its effectiveness.

Panel-MDP is a preliminary research on re-ranking within grid-based panels. However, the experiments were carried out in a restricted setting with small grids. To further substantiate its efficacy, additional baseline re-ranking methods could be incorporated. Future research will delve deeper into understanding user preferences in grid-based panels and will involve comparative experiments with a broader range of baseline methods on more intricate grids.

\begin{acks}
This work was funded by the National Natural Science Foundation of China (No. 62376275), Beijing Outstanding Young Scientist Program NO. BJJWZYJH012019100020098, the Fundamental Research Funds for the Central Universities, and the Research Funds of Renmin University of China (23XNKJ13).
This work was supported by Alibaba Group through Alibaba Innovative Research Program.
This work was funded by Intelligent Social Governance Interdisciplinary Platform, Major Innovation \& Planning Interdisciplinary Platform for the "Double-First Class" Initiative, Renmin University of China. The work was partially done at Beijing Key Laboratory of Big Data Management and Analysis Methods.
\end{acks}

\bibliographystyle{ACM-Reference-Format}
\balance
\bibliography{ref}


\begin{thebibliography}{24}


\ifx \showCODEN    \undefined \def \showCODEN     #1{\unskip}     \fi
\ifx \showDOI      \undefined \def \showDOI       #1{#1}\fi
\ifx \showISBNx    \undefined \def \showISBNx     #1{\unskip}     \fi
\ifx \showISBNxiii \undefined \def \showISBNxiii  #1{\unskip}     \fi
\ifx \showISSN     \undefined \def \showISSN      #1{\unskip}     \fi
\ifx \showLCCN     \undefined \def \showLCCN      #1{\unskip}     \fi
\ifx \shownote     \undefined \def \shownote      #1{#1}          \fi
\ifx \showarticletitle \undefined \def \showarticletitle #1{#1}   \fi
\ifx \showURL      \undefined \def \showURL       {\relax}        \fi
\providecommand\bibfield[2]{#2}
\providecommand\bibinfo[2]{#2}
\providecommand\natexlab[1]{#1}
\providecommand\showeprint[2][]{arXiv:#2}

\bibitem[Bello et~al\mbox{.}(2018)]%
        {bello2018seq2slate}
\bibfield{author}{\bibinfo{person}{Irwan Bello}, \bibinfo{person}{Sayali Kulkarni}, \bibinfo{person}{Sagar Jain}, \bibinfo{person}{Craig Boutilier}, \bibinfo{person}{Ed Chi}, \bibinfo{person}{Elad Eban}, \bibinfo{person}{Xiyang Luo}, \bibinfo{person}{Alan Mackey}, {and} \bibinfo{person}{Ofer Meshi}.} \bibinfo{year}{2018}\natexlab{}.
\newblock \showarticletitle{Seq2slate: Re-ranking and slate optimization with {RNN}s}.
\newblock \bibinfo{journal}{\emph{arXiv preprint arXiv:1810.02019}} (\bibinfo{year}{2018}).
\newblock


\bibitem[Cai et~al\mbox{.}(2018)]%
        {cai2018reinforcement}
\bibfield{author}{\bibinfo{person}{Qingpeng Cai}, \bibinfo{person}{Aris Filos-Ratsikas}, \bibinfo{person}{Pingzhong Tang}, {and} \bibinfo{person}{Yiwei Zhang}.} \bibinfo{year}{2018}\natexlab{}.
\newblock \showarticletitle{Reinforcement mechanism design for fraudulent behaviour in e-commerce}. In \bibinfo{booktitle}{\emph{Proceedings of the 32nd {AAAI} Conference on Artificial Intelligence}}. \bibinfo{pages}{957--964}.
\newblock


\bibitem[Chen et~al\mbox{.}(2023)]%
        {chen2023controllable}
\bibfield{author}{\bibinfo{person}{Sirui Chen}, \bibinfo{person}{Yuan Wang}, \bibinfo{person}{Zijing Wen}, \bibinfo{person}{Zhiyu Li}, \bibinfo{person}{Changshuo Zhang}, \bibinfo{person}{Xiao Zhang}, \bibinfo{person}{Quan Lin}, \bibinfo{person}{Cheng Zhu}, {and} \bibinfo{person}{Jun Xu}.} \bibinfo{year}{2023}\natexlab{}.
\newblock \showarticletitle{Controllable Multi-Objective Re-ranking with Policy Hypernetworks}. In \bibinfo{booktitle}{\emph{Proceedings of the 29th ACM SIGKDD Conference on Knowledge Discovery and Data Mining}}. \bibinfo{pages}{3855--3864}.
\newblock


\bibitem[Ge et~al\mbox{.}(2021)]%
        {ge2021towards}
\bibfield{author}{\bibinfo{person}{Yingqiang Ge}, \bibinfo{person}{Shuchang Liu}, \bibinfo{person}{Ruoyuan Gao}, \bibinfo{person}{Yikun Xian}, \bibinfo{person}{Yunqi Li}, \bibinfo{person}{Xiangyu Zhao}, \bibinfo{person}{Changhua Pei}, \bibinfo{person}{Fei Sun}, \bibinfo{person}{Junfeng Ge}, \bibinfo{person}{Wenwu Ou}, {et~al\mbox{.}}} \bibinfo{year}{2021}\natexlab{}.
\newblock \showarticletitle{Towards long-term fairness in recommendation}. In \bibinfo{booktitle}{\emph{Proceedings of the 14th ACM international conference on web search and data mining}}. \bibinfo{pages}{445--453}.
\newblock


\bibitem[Guo et~al\mbox{.}(2020)]%
        {guo2020debiasing}
\bibfield{author}{\bibinfo{person}{Ruocheng Guo}, \bibinfo{person}{Xiaoting Zhao}, \bibinfo{person}{Adam Henderson}, \bibinfo{person}{Liangjie Hong}, {and} \bibinfo{person}{Huan Liu}.} \bibinfo{year}{2020}\natexlab{}.
\newblock \showarticletitle{Debiasing grid-based product search in e-commerce}. In \bibinfo{booktitle}{\emph{Proceedings of the 26th ACM SIGKDD International Conference on Knowledge Discovery \& Data Mining}}. \bibinfo{pages}{2852--2860}.
\newblock


\bibitem[Huang et~al\mbox{.}(2021)]%
        {huang2021deep}
\bibfield{author}{\bibinfo{person}{Liwei Huang}, \bibinfo{person}{Mingsheng Fu}, \bibinfo{person}{Fan Li}, \bibinfo{person}{Hong Qu}, \bibinfo{person}{Yangjun Liu}, {and} \bibinfo{person}{Wenyu Chen}.} \bibinfo{year}{2021}\natexlab{}.
\newblock \showarticletitle{A deep reinforcement learning based long-term recommender system}.
\newblock \bibinfo{journal}{\emph{Knowledge-Based Systems}}  \bibinfo{volume}{213} (\bibinfo{year}{2021}), \bibinfo{pages}{106706}.
\newblock


\bibitem[Jiang et~al\mbox{.}(2017)]%
        {jiang2017learning}
\bibfield{author}{\bibinfo{person}{Zhengbao Jiang}, \bibinfo{person}{Ji-Rong Wen}, \bibinfo{person}{Zhicheng Dou}, \bibinfo{person}{Wayne~Xin Zhao}, \bibinfo{person}{Jian-Yun Nie}, {and} \bibinfo{person}{Ming Yue}.} \bibinfo{year}{2017}\natexlab{}.
\newblock \showarticletitle{Learning to diversify search results via subtopic attention}. In \bibinfo{booktitle}{\emph{Proceedings of the 40th international ACM SIGIR Conference on Research and Development in Information Retrieval}}. \bibinfo{pages}{545--554}.
\newblock


\bibitem[Kammerer and Gerjets(2010)]%
        {kammerer2010interface}
\bibfield{author}{\bibinfo{person}{Yvonne Kammerer} {and} \bibinfo{person}{Peter Gerjets}.} \bibinfo{year}{2010}\natexlab{}.
\newblock \showarticletitle{How the interface design influences users' spontaneous trustworthiness evaluations of web search results: Comparing a list and a grid interface}. In \bibinfo{booktitle}{\emph{Proceedings of the 2010 Symposium on Eye-Tracking Research \& Applications}}. \bibinfo{pages}{299--306}.
\newblock


\bibitem[Lillicrap et~al\mbox{.}(2015)]%
        {lillicrap2015continuous}
\bibfield{author}{\bibinfo{person}{Timothy~P Lillicrap}, \bibinfo{person}{Jonathan~J Hunt}, \bibinfo{person}{Alexander Pritzel}, \bibinfo{person}{Nicolas Heess}, \bibinfo{person}{Tom Erez}, \bibinfo{person}{Yuval Tassa}, \bibinfo{person}{David Silver}, {and} \bibinfo{person}{Daan Wierstra}.} \bibinfo{year}{2015}\natexlab{}.
\newblock \showarticletitle{Continuous control with deep reinforcement learning}.
\newblock \bibinfo{journal}{\emph{arXiv preprint arXiv:1509.02971}} (\bibinfo{year}{2015}).
\newblock


\bibitem[Liu et~al\mbox{.}(2023)]%
        {liu2023generative}
\bibfield{author}{\bibinfo{person}{Shuchang Liu}, \bibinfo{person}{Qingpeng Cai}, \bibinfo{person}{Zhankui He}, \bibinfo{person}{Bowen Sun}, \bibinfo{person}{Julian McAuley}, \bibinfo{person}{Dong Zheng}, \bibinfo{person}{Peng Jiang}, {and} \bibinfo{person}{Kun Gai}.} \bibinfo{year}{2023}\natexlab{}.
\newblock \showarticletitle{Generative Flow Network for Listwise Recommendation}. In \bibinfo{booktitle}{\emph{Proceedings of the 29th ACM SIGKDD Conference on Knowledge Discovery and Data Mining}}. \bibinfo{pages}{1524--1534}.
\newblock


\bibitem[Pei et~al\mbox{.}(2019)]%
        {pei2019personalized}
\bibfield{author}{\bibinfo{person}{Changhua Pei}, \bibinfo{person}{Yi Zhang}, \bibinfo{person}{Yongfeng Zhang}, \bibinfo{person}{Fei Sun}, \bibinfo{person}{Xiao Lin}, \bibinfo{person}{Hanxiao Sun}, \bibinfo{person}{Jian Wu}, \bibinfo{person}{Peng Jiang}, \bibinfo{person}{Junfeng Ge}, \bibinfo{person}{Wenwu Ou}, {et~al\mbox{.}}} \bibinfo{year}{2019}\natexlab{}.
\newblock \showarticletitle{Personalized re-ranking for recommendation}. In \bibinfo{booktitle}{\emph{Proceedings of the 13th ACM conference on recommender systems}}. \bibinfo{pages}{3--11}.
\newblock


\bibitem[Schulman et~al\mbox{.}(2015)]%
        {schulman2015high}
\bibfield{author}{\bibinfo{person}{John Schulman}, \bibinfo{person}{Philipp Moritz}, \bibinfo{person}{Sergey Levine}, \bibinfo{person}{Michael Jordan}, {and} \bibinfo{person}{Pieter Abbeel}.} \bibinfo{year}{2015}\natexlab{}.
\newblock \showarticletitle{High-dimensional continuous control using generalized advantage estimation}.
\newblock \bibinfo{journal}{\emph{arXiv preprint arXiv:1506.02438}} (\bibinfo{year}{2015}).
\newblock


\bibitem[Schulman et~al\mbox{.}(2017)]%
        {schulman2017proximal}
\bibfield{author}{\bibinfo{person}{John Schulman}, \bibinfo{person}{Filip Wolski}, \bibinfo{person}{Prafulla Dhariwal}, \bibinfo{person}{Alec Radford}, {and} \bibinfo{person}{Oleg Klimov}.} \bibinfo{year}{2017}\natexlab{}.
\newblock \showarticletitle{Proximal policy optimization algorithms}.
\newblock \bibinfo{journal}{\emph{arXiv preprint arXiv:1707.06347}} (\bibinfo{year}{2017}).
\newblock


\bibitem[Shani et~al\mbox{.}(2005)]%
        {shani2005mdp}
\bibfield{author}{\bibinfo{person}{Guy Shani}, \bibinfo{person}{David Heckerman}, \bibinfo{person}{Ronen~I Brafman}, {and} \bibinfo{person}{Craig Boutilier}.} \bibinfo{year}{2005}\natexlab{}.
\newblock \showarticletitle{An MDP-based recommender system}.
\newblock \bibinfo{journal}{\emph{Journal of Machine Learning Research}} \bibinfo{volume}{6}, \bibinfo{number}{9} (\bibinfo{year}{2005}), \bibinfo{pages}{1265--1295}.
\newblock


\bibitem[Verstrepen and Goethals(2015)]%
        {verstrepen2015top}
\bibfield{author}{\bibinfo{person}{Koen Verstrepen} {and} \bibinfo{person}{Bart Goethals}.} \bibinfo{year}{2015}\natexlab{}.
\newblock \showarticletitle{Top-n recommendation for shared accounts}. In \bibinfo{booktitle}{\emph{Proceedings of the 9th ACM Conference on Recommender Systems}}. \bibinfo{pages}{59--66}.
\newblock


\bibitem[Wang et~al\mbox{.}(2019)]%
        {wang2019sequential}
\bibfield{author}{\bibinfo{person}{Fan Wang}, \bibinfo{person}{Xiaomin Fang}, \bibinfo{person}{Lihang Liu}, \bibinfo{person}{Yaxue Chen}, \bibinfo{person}{Jiucheng Tao}, \bibinfo{person}{Zhiming Peng}, \bibinfo{person}{Cihang Jin}, {and} \bibinfo{person}{Hao Tian}.} \bibinfo{year}{2019}\natexlab{}.
\newblock \showarticletitle{Sequential evaluation and generation framework for combinatorial recommender system}.
\newblock \bibinfo{journal}{\emph{arXiv preprint arXiv:1902.00245}} (\bibinfo{year}{2019}).
\newblock


\bibitem[Xie et~al\mbox{.}(2017)]%
        {xie2017investigating}
\bibfield{author}{\bibinfo{person}{Xiaohui Xie}, \bibinfo{person}{Yiqun Liu}, \bibinfo{person}{Xiaochuan Wang}, \bibinfo{person}{Meng Wang}, \bibinfo{person}{Zhijing Wu}, \bibinfo{person}{Yingying Wu}, \bibinfo{person}{Min Zhang}, {and} \bibinfo{person}{Shaoping Ma}.} \bibinfo{year}{2017}\natexlab{}.
\newblock \showarticletitle{Investigating examination behavior of image search users}. In \bibinfo{booktitle}{\emph{Proceedings of the 40th International ACM SIGIR Conference on Research and Development in Information Retrieval}}. \bibinfo{pages}{275--284}.
\newblock


\bibitem[Xie et~al\mbox{.}(2019)]%
        {xie2019grid}
\bibfield{author}{\bibinfo{person}{Xiaohui Xie}, \bibinfo{person}{Jiaxin Mao}, \bibinfo{person}{Yiqun Liu}, \bibinfo{person}{Maarten de Rijke}, \bibinfo{person}{Yunqiu Shao}, \bibinfo{person}{Zixin Ye}, \bibinfo{person}{Min Zhang}, {and} \bibinfo{person}{Shaoping Ma}.} \bibinfo{year}{2019}\natexlab{}.
\newblock \showarticletitle{Grid-based evaluation metrics for web image search}. In \bibinfo{booktitle}{\emph{Proceedings of the 2019 World Wide Web Conference}}. \bibinfo{pages}{2103--2114}.
\newblock


\bibitem[Xu et~al\mbox{.}(2020)]%
        {xu2020reinforcement}
\bibfield{author}{\bibinfo{person}{Jun Xu}, \bibinfo{person}{Zeng Wei}, \bibinfo{person}{Long Xia}, \bibinfo{person}{Yanyan Lan}, \bibinfo{person}{Dawei Yin}, \bibinfo{person}{Xueqi Cheng}, {and} \bibinfo{person}{Ji-Rong Wen}.} \bibinfo{year}{2020}\natexlab{}.
\newblock \showarticletitle{Reinforcement learning to rank with pairwise policy gradient}. In \bibinfo{booktitle}{\emph{Proceedings of the 43rd International ACM SIGIR Conference on Research and Development in Information Retrieval}}. \bibinfo{pages}{509--518}.
\newblock


\bibitem[Xue et~al\mbox{.}(2023)]%
        {xue2023prefrec}
\bibfield{author}{\bibinfo{person}{Wanqi Xue}, \bibinfo{person}{Qingpeng Cai}, \bibinfo{person}{Zhenghai Xue}, \bibinfo{person}{Shuo Sun}, \bibinfo{person}{Shuchang Liu}, \bibinfo{person}{Dong Zheng}, \bibinfo{person}{Peng Jiang}, \bibinfo{person}{Kun Gai}, {and} \bibinfo{person}{Bo An}.} \bibinfo{year}{2023}\natexlab{}.
\newblock \showarticletitle{PrefRec: Recommender Systems with Human Preferences for Reinforcing Long-term User Engagement}.
\newblock  (\bibinfo{year}{2023}).
\newblock


\bibitem[Zhao et~al\mbox{.}(2018)]%
        {zhao2018deep}
\bibfield{author}{\bibinfo{person}{Xiangyu Zhao}, \bibinfo{person}{Long Xia}, \bibinfo{person}{Liang Zhang}, \bibinfo{person}{Zhuoye Ding}, \bibinfo{person}{Dawei Yin}, {and} \bibinfo{person}{Jiliang Tang}.} \bibinfo{year}{2018}\natexlab{}.
\newblock \showarticletitle{Deep reinforcement learning for page-wise recommendations}. In \bibinfo{booktitle}{\emph{Proceedings of the 12th ACM Conference on Recommender Systems}}. \bibinfo{pages}{95--103}.
\newblock


\bibitem[Zhuang et~al\mbox{.}(2018)]%
        {zhuang2018globally}
\bibfield{author}{\bibinfo{person}{Tao Zhuang}, \bibinfo{person}{Wenwu Ou}, {and} \bibinfo{person}{Zhirong Wang}.} \bibinfo{year}{2018}\natexlab{}.
\newblock \showarticletitle{Globally optimized mutual influence aware ranking in e-commerce search}.
\newblock \bibinfo{journal}{\emph{arXiv preprint arXiv:1805.08524}} (\bibinfo{year}{2018}).
\newblock


\bibitem[Zou et~al\mbox{.}(2019)]%
        {zou2019reinforcement}
\bibfield{author}{\bibinfo{person}{Lixin Zou}, \bibinfo{person}{Long Xia}, \bibinfo{person}{Zhuoye Ding}, \bibinfo{person}{Jiaxing Song}, \bibinfo{person}{Weidong Liu}, {and} \bibinfo{person}{Dawei Yin}.} \bibinfo{year}{2019}\natexlab{}.
\newblock \showarticletitle{Reinforcement learning to optimize long-term user engagement in recommender systems}. In \bibinfo{booktitle}{\emph{Proceedings of the 25th ACM SIGKDD International Conference on Knowledge Discovery \& Data Mining}}. \bibinfo{pages}{2810--2818}.
\newblock


\bibitem[Zou et~al\mbox{.}(2020)]%
        {zou2020pseudo}
\bibfield{author}{\bibinfo{person}{Lixin Zou}, \bibinfo{person}{Long Xia}, \bibinfo{person}{Pan Du}, \bibinfo{person}{Zhuo Zhang}, \bibinfo{person}{Ting Bai}, \bibinfo{person}{Weidong Liu}, \bibinfo{person}{Jian-Yun Nie}, {and} \bibinfo{person}{Dawei Yin}.} \bibinfo{year}{2020}\natexlab{}.
\newblock \showarticletitle{Pseudo Dyna-Q: A reinforcement learning framework for interactive recommendation}. In \bibinfo{booktitle}{\emph{Proceedings of the 13th International Conference on Web Search and Data Mining}}. \bibinfo{pages}{816--824}.
\newblock


\end{thebibliography}

\end{document}